\documentclass[aps,superscriptaddress,eqsecnum,nofootinbib,showpacs,preprintnumbers]
{revtex4}
\usepackage{graphicx,epsfig}
\usepackage{amsmath}
\usepackage {amssymb}

\newcommand{\be}{\begin{eqnarray}}
\newcommand{\ee}{\end{eqnarray}}
\newcommand{\bea}{\begin{eqnarray}}

\newcommand{\eea}{\end{eqnarray}}

\makeindex

\begin{document}

\title{Fermionic field perturbations of a three-dimensional Lifshitz black hole in conformal gravity}

\author{P. A. Gonz\'{a}lez}
\email{pablo.gonzalez@udp.cl} \affiliation{Facultad de
Ingenier\'{i}a y Ciencias, Universidad Diego Portales, Avenida Ej\'{e}rcito
Libertador 441, Casilla 298-V, Santiago, Chile.}

\author{Yerko V\'asquez}
\email{yvasquez@userena.cl}
\affiliation{Departamento de F\'isica y Astronom\'ia, Facultad de Ciencias, Universidad de La Serena,\\
Avenida Cisternas 1200, La Serena, Chile.}

\author{Ruth Noemi Villalobos}
\email{noemi@dfuls.cl}
\affiliation{Departamento de F\'isica y Astronom\'ia, Facultad de Ciencias, Universidad de La Serena,\\
Avenida Cisternas 1200, La Serena, Chile.}

\date{\today }

\begin{abstract}
We study the propagation of massless fermionic fields in the background of a three-dimensional Lifshitz black hole, which is a solution of conformal gravity. The black hole solution is characterized by a null dynamical exponent. Then, we compute analytically the quasinormal modes, the area spectrum, and the absorption cross section for fermionic fields. The analysis of the quasinormal modes shows that the fermionic perturbations are stable in this background. The area and entropy spectrum are evenly spaced. At the low frequency limit, it is observed that there is a range of values of the angular momentum of the mode that contributes to the absorption cross section, whereas it vanishes at the high frequency limit. In addition, by a suitable change of variables a gravitational soliton can also be obtained and the stability of the quasinormal modes are studied and ensured.

\end{abstract}

\maketitle
\newpage
\tableofcontents

%\vspace{5cm}
\newpage
\section{Introduction}

The three-dimensional models of gravity have attracted remarkable interest in the last decade. Apart from three-dimensional general relativity (GR), an interesting model of gravity is described by topologically massive gravity (TMG), which modifies GR by adding a Chern-Simons gravitational term to the action \cite{Deser:1981wh}. In contrast to GR in 3 dimensions, TMG has a propagating degree of freedom, which corresponds to a massive graviton. Also, another important characteristic of TMG is the possibility of constructing a chiral theory of gravity at a special point of the space of parameters \cite{Deser:1982vy}. Another three-dimensional model that has received considerable attention is new massive gravity (NMG), where the action is the standard Einstein-Hilbert term plus a specific combination of scalar curvature square term and a Ricci tensor square term \cite{Bergshoeff:2009hq, Bergshoeff:2009tb, Andringa:2009yc}, and at the linearized level it is equivalent to the Fierz-Pauli action for a massive spin-2 field \cite{Bergshoeff:2009hq}. Solutions of TMG and NMG and further aspects can be found in Refs. \cite{Garbarz:2008qn, Clement:2009gq, AyonBeato:2009yq, Clement:2009ka, AyonBeato:2009nh, Oliva:2009ip, Correa:2014ika, Nakasone:2009bn, Bergshoeff:2009aq, Oda:2009ys, Deser:2009hb} and references therein. TMG and NMG share common features; however, there are also different aspects: one of these is the existence in NMG of black holes known as new type black holes. These solutions also appear in  the three-dimensional conformal gravity \cite{Oliva:2009hz}. The action of three-dimensional conformal gravity in vacuum is just given by the Chern-Simons gravitational term, and the equations of motion contain third derivatives of the metric. Some solutions of this model have been studied in \cite{Oliva:2009hz, Guralnik:2003we, Grumiller:2003ad}. Additionally, in \cite{Catalan:2014una} a Lifshitz black hole solution to this theory with $z=0$ was obtained. On the other hand, in four spacetime dimensions, conformal gravity is a four-derivative theory and is perturbatively renormalizable \cite{Stelle:1976gc, Stelle:1977ry}. Other Lifshitz black hole solutions have been found in \cite{Lu:2012xu} in four-dimensional conformal gravity and in \cite{Lu:2013hx} in six-dimensional conformal gravity.

The Lifshitz spacetimes present an anisotropic scale invariance $t\rightarrow \lambda ^{z}t$, $x^i\rightarrow \lambda x^i$ along with a scaling $r\rightarrow \lambda^{-1} r$ for the radial coordinate, where the dynamical exponent $z$ is the relative scale dimension of time $t$ and space $x^{i}$ coordinates. These spacetimes are important in the study of dual field theories with anisotropic scale invariance and are described by the following metrics \cite{Kachru:2008yh}:
\begin{equation}
ds^2=- \frac{r^{2z}}{\ell^{2z}}dt^2+\frac{\ell^2}{r^2}dr^2+\frac{r^2}{\ell^2} d\vec{x}%
^2~,  \label{lif1}
\end{equation}
where $\vec{x}$ represents a $D-2$ dimensional spatial vector, $D$ is the
spacetime dimension and $\ell$ denotes the length scale in these geometries. It is worth mentioning that for $z=0$ the previously mentioned anisotropic scale invariance corresponds to space-like scale invariance with no transformation of time.
Several black holes whose asymptotic behavior is described by (\ref{lif1}) (Lifshitz black holes) have been reported, see for instance Refs. \cite{AyonBeato:2009nh, Solutions}.  Also, by introducing both an Abelian gauge field and a scalar dilaton, spacetimes emerge which, in addition to having an anisotropic scaling exponent $z$ as the Lifshitz metric, have an overall hyperscaling violating factor with hyperscaling exponent {\bf{$\eta$}} that is not scale invariant; thus, this line element is conformally related to the Lifshitz metric and transforms as $ds\rightarrow \lambda ^{\eta/(D-2)}ds$ under the Lifshitz scaling. This spacetime is important in the study of dual field theories with hyperscaling violation  \cite{Dong:2012se, Narayan:2012hk, Perlmutter:2012he, Ammon:2012je, Bhattacharya:2012zu, Dey:2012fi, Alishahiha:2012qu, Gath:2012pg, Bueno:2012vx, Iizuka:2012pn, Fan:2013tpa, Fan:2013zqa, Fan:2013tga, Bhattacharya:2014dea, Ogawa:2011bz, Huijse:2011ef}, and were also investigated in other gravitational theories \cite{Dehghani:2015gza, Ganjali:2015cba, Feng:2015yja}.

One important characteristic of black holes is their quasinormal modes (QNMs), which nowadays are of great interest due to the observation of gravitational waves from the merger of two black holes \cite{Abbott:2016blz}. Nevertheless, the observed signal is consistent with the Einstein gravity \cite{TheLIGOScientific:2016src}, the window for alternative theories is also open \cite{Konoplya:2016pmh} mainly owing to large uncertainties in mass and angular momenta of the ringing black hole. However, the QNMs and their quasinormal frequencies (QNFs)
have a long history \cite{Regge:1957td, Zerilli:1971wd, Zerilli:1970se,
Kokkotas:1999bd, Nollert:1999ji, Konoplya:2011qq}. The QNMs provide information about the stability of matter fields that evolve perturbatively in the exterior region of the black holes, without backreacting on the metric. Also, in the context of AdS/CFT correspondence \cite{Maldacena:1997re}, the QNMs determine how fast a thermal
state in the boundary theory will reach thermal equilibrium \cite{Horowitz:1999jd}. 
In addition,
in the context of black hole thermodynamics, the
QNMs allow us to study the quantum area spectrum of the black hole horizon 
as well as the mass and the entropy spectrum. Moreover, in \cite{Corda:2012tz} the authors
discuss a connection between Hawking radiation and black hole quasinormal
modes, which is important on the route to quantize gravity, because one can
naturally interpret black hole quasinormal modes in terms of quantum levels.
On the other hand, the Hawking radiation emitted at the event horizon may be modified as this will no longer be that of a black body, when an observer located very far away from the black hole measures the spectrum  \cite{Maldacena:1996ix}. The factors that modify the spectrum emitted by a black hole are known as greybody factors and these can be obtained through classical scattering; their study therefore allows the semiclassical gravity dictionary to be increased, and also offers insight into the quantum nature of black holes and thus of quantum gravity; for a review of this topic see \cite{Harmark:2007jy}.

The aim of this work is to study the propagation of fermionic fields in the background of a Lifshitz black hole with $z=0$ in three-dimensional conformal gravity \cite{Catalan:2014una}, and to study the stability of the fermionic field in this background by computing the exact QNMs, to compute the area spectrum, and to analyze the greybody factor. Exact solutions for the QNMs of black holes in $2+1$ -dimensional spacetimes can be found in \cite{exactQNM}, \cite{Catalan:2014una}.

The QNFs have been calculated by means of numerical and analytical techniques; 
the QNMs of Lifshitz black holes under scalar field perturbations have been studied in \citep{CuadrosMelgar:2011up, Gonzalez:2012de, Gonzalez:2012xc, Myung:2012cb, Becar:2012bj,Giacomini:2012hg, Lepe:2012zf, Catalan:2014ama, Catalan:2014una, Sybesma:2015oha, Becar:2015kpa, Zangeneh:2017rhc}, and generally the scalar modes of Lifshitz black holes are stable. Non-relativistic fermion Green's functions in 4-dimensional Lifshitz spacetime with $z=2$ were studied in \cite{Alishahiha:2012nm} by considering fermions in this background and a non-relativistic (mixed) boundary condition, and it was shown that the Green's functions have a flat band.
Also, the Dirac QNMs of a 4-dimensional Lifshitz black hole were studied in \cite{Catalan:2013eza} and the electromagnetic QNMs in \cite{Lopez-Ortega:2014oha}. In addition, QNMs of Lifshitz black holes with hyperscaling violation have been considered in Refs. \cite{BAI:2013koa, Gonzalez:2015gla, Becar:2015gca}.
On the other hand, 
the area spectrum of three-dimensional Lifshitz black holes
was studied in \cite{CuadrosMelgar:2011up} and in \cite{Fernando:2015gta},  which turns out to be equally spaced.
Additionally, the scalar greybody factors for an asymptotically Lifshitz black hole were studied in \cite{Gonzalez:2012xc, Lepe:2012zf}.
 
The paper is organized as follows: In Sec. II we give a brief review of a three-dimensional Lifshitz black hole with dynamical exponent $z=0$ in conformal gravity. Also, by means of a suitable change of variable, a gravitational soliton solution is obtained.  Then, in Sec. III we solve analytically the Dirac equation in this background and we find the QNMs, the area spectrum, the reflection and transmission coefficients, and the absorption cross section. We conclude with final remarks in Sec. IV.

\section{Lifshitz black hole and gravitational soliton in three-dimensional conformal gravity}
The field equations of three-dimensional conformal gravity in vacuum are given by the vanishing of the Cotton tensor:
\begin{equation} \label{fieldeq}
C^{\alpha}_{\,\,\beta}=\epsilon ^{\rho \sigma \alpha} \nabla_{\rho} \left( R_{\sigma \beta}-\frac{1}{4}g_{\sigma \beta} R\right)=0~,
\end{equation}
where $R_{\alpha \beta}$ and $R$ denote the Ricci tensor and the Ricci scalar, respectively. The Cotton tensor is a traceless tensor that vanishes if and only if the metric is locally conformally flat. Interesting solutions to this theory have been studied, for instance, in \cite{Guralnik:2003we, Grumiller:2003ad, Oliva:2009hz, Catalan:2014una}. The black hole background that we will consider is given by the following solution of (\ref{fieldeq}) obtained in \cite{Catalan:2014una}:
\begin{eqnarray} \label{solution}
ds^{2} & = & - f(r )dt^{2}+\frac{\ell^2}{r^{2}}\frac{dr^{2}}{f ( r )}+r^{2}d\phi ^{2}~, \\
f ( r ) & = & 1-\frac{r_{+}^2}{r^{2}}~.
\end{eqnarray}
The above metric describes an asymptotically Lifshitz black hole with $z=0$. The event horizon is located at $r=r_{+}$, where $r_{+}$ is an integration constant related to the mass of the circular object. 
Interestingly, the Lifshitz spacetime for $z=0$, which is a solution of conformal gravity and to which metric (\ref{solution}) tends asymptotically:
\begin{equation}
ds^2 \Big| _{r \rightarrow \infty} \sim-dt^2+\frac{\ell^2}{r^2}dr^2+r^2 d \phi^2~,
\end{equation}
can be written, by means of the double Wick rotation $t \rightarrow -i \ell \tilde{\phi}$ and $\phi \rightarrow i \tilde{t}$ and along with the coordinate transformation $\tilde{r}=\frac{\ell^2}{r}$, as $AdS_2 \times S^1$
\begin{equation}\label{ads}
ds^2=\frac{\ell^2}{\tilde{r}^2} \left( -d \tilde{t}^2+d \tilde{r}^2 \right) +\ell^2 d \tilde{\phi} ^2~,
\end{equation}
where $\ell$ corresponds to the length scale of $AdS_2$. In addition, by performing a double Wick rotation $t \rightarrow - i \ell \tilde{\phi}$ and $\phi \rightarrow i \tilde{t}/r_{+}$ to the solution (\ref{solution}) along with the coordinate transformation $r=r_{+} \cosh ( \tilde{r} /\ell)$, the following  gravitational soliton solution can be found:
\begin{equation}\label{gso}
ds^2 = \tanh ^2 ( \tilde{r}/ \ell ) \ell^2 d \tilde{\phi}^2 + d \tilde{r}^2- \cosh ^2 ( \tilde{r} / \ell ) d \tilde{t}^2~.
\end{equation}
The Kretschmann scalar of this soliton is regular everywhere and is given by 
\begin{equation}
R_{\mu \nu \rho \sigma }R^{\mu \nu \rho \sigma }=\frac{4}{\ell^4} \left( 1+ 5 \text{sech} ^4  \left( {\frac{\tilde{r}}{\ell}}\right) \right)~.
\end{equation}
The soliton tends asymptotically to $AdS_2 \times S^1$ (\ref{ads}), which can be verified by means of the change of variable $\rho=\frac{\ell}{\cosh (\tilde{r}/\ell)}$.
As we have shown, three dimensional conformal gravity in vacuum admits nontrivial solutions, such as the Lifshitz black hole (\ref{solution}) and the gravitational soliton (\ref{gso}). Gravitational soliton solutions to NMG have been reported in \cite{Oliva:2009ip, Gonzalez:2011nz}. In the following we will set $\ell=1$.

The temperature of the Lifshitz black hole can be found by evaluating the surface gravity, which is defined by $\kappa^2=-\frac{1}{2}(\nabla_{\mu} \chi_{\nu})(\nabla^{\mu} \chi^{\nu})$, where the time-like Killing vector is $\chi^{\nu}=(1,0,0)$; So, $\kappa=\frac{1}{2}r_+ f'(r_+)=1$ and the Hawking temperature is found to be a constant,
\begin{equation}
T=\frac{\kappa}{2 \pi}=\frac{1}{2 \pi}~.
\end{equation}
On the other hand, the ADM mass $\mathcal{M}$ of this black hole has been obtained in \cite{Fernando:2015gta} and is given by $\mathcal{M}=r_{+}$. In order to determine here the conserved quantities we will employ the method developed by Padmanabhan in \cite{Padmanabhan:2012gx}, which is based on the relation between the first law of thermodynamics and the equations of motion, and it seems to be applicable to a very wide class of theories. In this approach the conserved quantities can be obtained directly from the equations of motion. In our case, the only nonvanishing component of the field equations comes from the non-null component of the Cotton tensor $C_{t \phi}$=$C_{\phi t}$, and it is given by
\begin{equation}
-3f'(r)+r(3f''(r)+rf'''(r))=0~.
\end{equation}
Evaluating the second and third derivatives of $f(r)=1-r_{+}^2/r^2$ at the horizon $r_+$ in the above equation we get:
\begin{equation}
r_+ f'(r_+)=2~, 
\end{equation}
so, rewriting this equation in the form of the first law of thermodynamics, by multiplying it by the factor $\frac{1}{2} dr_+$, we obtain
\begin{equation}
d r_+= \frac{r_+  f'(r_+)}{4 \pi} d(2 \pi r_+)~. 
\end{equation}
From this equation, identifying the Hawking temperature $T=\frac{r_+  f'(r_+)}{4 \pi}=\frac{1}{2 \pi}$, we obtain that the mass is given by $\mathcal{M}=r_+$ and the entropy $S=2 \pi r_+$, which is equal to the horizon area.

\section{Fermionic perturbations}
\label{FP}
In this section we will consider a matter distribution described by a fermionic field in the probe limit outside the event horizon of the black hole under consideration (\ref{solution}). 
It is worth mentioning that the Cotton tensor is traceless and this implies that the trace of the stress-energy tensor of the matter fields must be traceless, too. However, for a probe field it is not actually necessary to respect the same symmetries of the Cotton tensor; however, if one goes beyond the probe-field approximation, this must be respected, and therefore we will consider a massless fermionic field which has a traceless stress-energy tensor. The fermionic perturbations in this background are governed by the Dirac equation in curved spacetime
\begin{equation}\label{DE}
\gamma ^{\mu }\nabla _{\mu } \psi =0~,
\end{equation}%
where the covariant derivative is defined as 
\begin{equation}
\nabla _{\mu }=\partial _{\mu }+\frac{1}{2}\omega _{\text{ \ \ \ }\mu
}^{ab}J_{ab}~,
\end{equation}%
and the generators of the Lorentz group $J_{ab}$ are given by $J_{ab}=\frac{1}{4}\left[ \gamma _{a},\gamma _{b}\right]$.
The gamma matrices in curved spacetime $\gamma ^{\mu }$ are defined by $\gamma ^{\mu }=e_{\text{ \ }a}^{\mu }\gamma ^{a}$,
where $\gamma ^{a}$ are the gamma matrices in flat spacetime. In order to
solve the Dirac equation we use the diagonal vielbein 
\begin{equation}
e^{0}=\sqrt{f} dt~, \,\,\,\,\, e^{1}=\frac{1}{r \sqrt{f}}dr~, \,\,\,\,\, e^{2}=r d \phi ~.
\end{equation}
The spin connection can be obtained directly from the null torsion condition $de^{a}+\omega _{\text{ \ }b}^{a}e^{b}=0$,
and its non null components are
\begin{equation}
\omega ^{01}=\frac{r f^{\prime}\left( r\right)}{2} dt~, \,\,\,\,\, \omega ^{12}=-r\sqrt{f} d \phi ~.
\end{equation}
Now, by using
the following representation of the gamma matrices $\gamma ^{0}=i\sigma ^{2}$, $\gamma ^{1}=\sigma^{1}$, and $\gamma^2=\sigma^3$,
where $\sigma ^{i}$ are the Pauli matrices, 
along with the
following ansatz for the fermionic field
\begin{equation}
\psi =e^{-i\omega t} e^{i \kappa \phi }\frac{1}{r^{1/2}f^{1/4}}  \left( 
\begin{array}{c}
\psi _{1} \\ 
\psi _{2}
\end{array}
\right)~,
\end{equation}
we obtain the following set of differential equations
\begin{eqnarray}
\label{system1}r\sqrt{f}\partial _{r}\psi _{1}+\frac{i\omega}{\sqrt{f}}\psi _{1}-\frac{i \kappa}{r}\psi _{2}&=&0~, \notag \\
\label{system}r\sqrt{f}\partial _{r}\psi _{2}-\frac{i\omega}{\sqrt{f}}\psi _{2}+\frac{i \kappa }{ r }\psi _{1}&=&0~,
\end{eqnarray}
where $\kappa$ corresponds to the angular momentum of the mode and can take only integer values. Decoupling the above system of equations, we obtain the following equations for $\psi_1$ and $\psi_2$
\begin{eqnarray}
\label{radial1}&& \psi_1''(r)+\left( \frac{2}{r} +\frac{f'(r)}{2f(r)}\right) \psi_1'(r)+\frac{1}{2r^4f(r)^2} \left( -2 f(r)(\kappa^2-i \omega r^2) +\omega r^2 (2 \omega-i r f'(r)) \right) \psi_1=0 \,~, \\
\label{radial2}&& \psi_2''(r)+\left( \frac{2}{r} +\frac{f'(r)}{2f(r)}\right) \psi_2'(r)+\frac{1}{2r^4f(r)^2} \left( -2 f(r)(\kappa^2+i \omega r^2) +\omega r^2 (2 \omega+i r f'(r)) \right) \psi_2=0 \,~.
\end{eqnarray}
\subsection{Stability analysis}
\begin{figure}[!h]
 \begin{center}
  \includegraphics[width=0.60\textwidth]{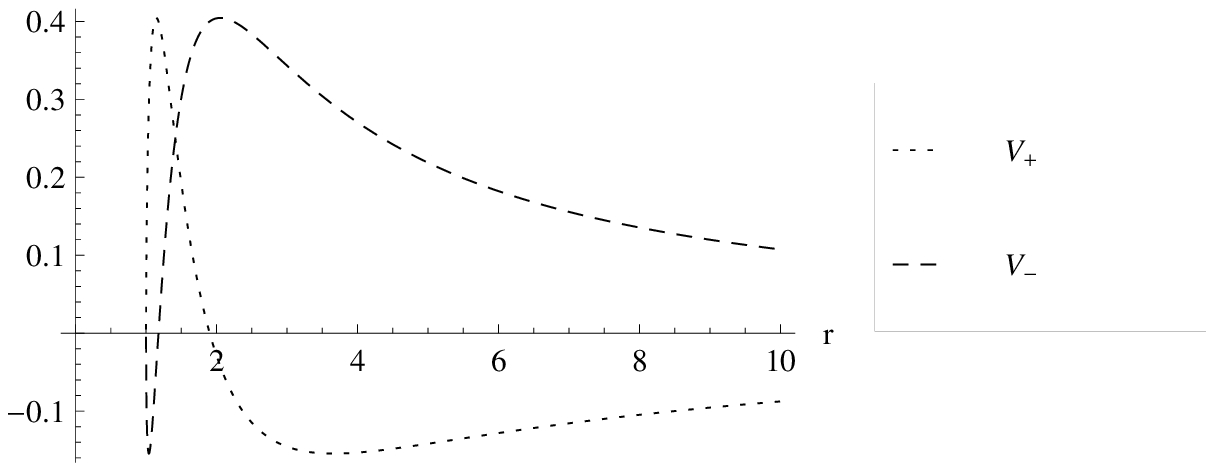}
 \end{center}
 \caption{The behavior of the effective potentials $V_{+}$ (dotted line) and $V_{-}$ (dashed line) for $r_{+}=1$ and $\kappa=1$}
 \label{F1}
\end{figure}

Defining $Z_{\pm}=\psi_1 \pm i \psi_2$ and $W=\kappa \sqrt{f}/r$, see \cite{Chandrasekhar}, and performing a change of variable to the tortoise coordinate $x=(1/2) \ln (r^2-r_{+}^2)$, where we have defined $dx=dr/(r f(r))$, the following differential equations are obtained:
\begin{eqnarray}
\label{a1}&& -i \omega Z_{+}-\partial_x Z_{-}= W Z_{-}~,  \\
\label{a2}&& -i \omega Z_{-}-\partial_x Z_{+}= -W Z_{+}~,
\end{eqnarray}
from the coupled system of equations (\ref{system1}). Now, decoupling (\ref{a1}) and (\ref{a2}), we obtain the following Schr\"{o}dinger-like equations:
\begin{equation}\label{Schrodinger}
(-\partial^{2}_{x}+V_{\pm}) Z_{\pm}=\omega^2 Z_{\pm}~,
\end{equation}
where the effective potentials are given by
\begin{equation}
V_{\pm}=W^2\pm \partial_{x} W=\kappa f \left(\frac{1}{r^2}\mp \frac{\sqrt{f}}{r} \pm \frac{f' }{2 \sqrt{f}} \right)~.
\end{equation}
In terms of the tortoise coordinate, the function $W(x)$ and the potentials $V_{\pm}(x)$ read
\begin{equation}
W(x)=\frac{\kappa e^x}{e^{2x}+r_{+}^2}~, \,\,\,\,\, V_{\pm}(x)=\mp \frac{\kappa e^x ( e^{2x}\mp\kappa e^x -r_{+}^2)}{(e^{2x}+r_{+}^2)^2}~.\label{W}
\end{equation}
In Fig. \ref{F1} we plot the effective potentials for $r_{+}=1$ and $\kappa=1$. We can observe that the potentials are not positive-definite and they are zero at the horizon and at spatial infinity.
Therefore, suitable boundary conditions for the fermionic field are purely an ingoing wave at the event horizon and an outgoing wave at spatial infinity:
\begin{equation}\label{bc}
Z_{\pm}\big|_{x \rightarrow -\infty} \propto e^{-i \omega x}~, \,\,\,\,\, Z_{\pm}\big|_{x \rightarrow \infty} \propto e^{i \omega x}~.
\end{equation}
Now, we shall show that the classical stability of the massless fermionic field in this background can be proven using the S-deformation method \cite{Kodama:2003ck}. So, multiplying Eq. (\ref{Schrodinger}) by $Z_{\pm}^*$, then performing an integration by parts and taking into account the boundary conditions (\ref{bc}), one can arrive at the following expression:
\begin{equation}\label{condition}
\int_{-\infty}^{\infty} \left( V_{\pm}(r) |Z_{\pm}(x)|^2+\left| \frac{dZ_{\pm}(x)}{dx}\right|^2 \right) dx= i \omega (|Z_{\pm}(x=\infty)|^2+|Z_{\pm}(x=-\infty)|^2)+\omega^2 \int_{-\infty}^{\infty} |Z_{\pm}(x)|^2 dx~.
\end{equation}
From this expression one can conclude that the imaginary part of $\omega$ is always negative when $V_{\pm}(r)>0$ in the region outside the event horizon, see \cite{Kodama:2003ck} and \cite{Zhidenko:2009zx} for details. The potentials $V_{\pm}$ are not positive-definite; however, in the S-deformation method a new derivative $D=\frac{d}{dx}+S(x)$ is defined, and the first integral in (\ref{condition}) can be written as
\begin{equation}
\int_{-\infty}^{\infty} \left( V_{\pm}(r) |Z_{\pm}(x)|^2+\left| \frac{dZ_{\pm}(x)}{dx}\right|^2 \right) dx=\int_{-\infty}^{\infty}\left( \tilde{V}_{\pm} |Z_{\pm}(x)|^2+|D Z_{\pm}|^2 \right) dx-S(x) |Z_{\pm}(x)|^2 \Bigg| _{x=-\infty}^{x=\infty}~,
\end{equation}
where $\tilde{V}_{\pm}=V_{\pm}+\frac{dS}{dx}-S^2$. Appropriate functions are given by $S=-W$ for $V_{+}$ and $S=W$ for $V_{-}$ \cite{Zhidenko:2009zx}, \cite{LopezOrtega:2012hx}. In this case, $\tilde{V}_{\pm}=0$ and from (\ref{W}) we obtain $W(x=-\infty) = 0$ and $W(x=\infty) = 0$, which ensures that the integral is positive, and thus guarantees the stability of the fermionic field. Having demonstrated that the propagation of massless fermionic fields is stable in this background, in the following we will compute analytically the QNMs in order to study their behavior and to determine the area spectrum.

\subsection{Quasinormal modes}

Under the change of variable $z=1-r_{+}^2/r^2$, the radial equation \eqref{radial1} can be written as
\begin{equation}
z(1-z)\psi_1''(z)+(1/2-z) \psi_1'(z)+(1/4) ( -\kappa^2/M+i\omega /(1-z)+\omega^2/(z(1-z))-i\omega /z ) \psi_1(z)=0~,\
\end{equation}
and if in addition we define $\psi_1(z)= z^{\alpha} (1-z)^{\beta} F (z)$, the above equation leads to the hypergeometric equation
\begin{equation}\label{hyp}
z\left( 1-z\right) F^{\prime \prime }\left( z\right) +\left( c-\left(
1+a+b\right) z\right) F^{\prime }\left( z\right) -abF\left( z\right) =0~, 
\end{equation}
where
\begin{eqnarray}
\alpha_+&=&\frac{1}{2}+\frac{i \omega}{2}~, \, \alpha_-=-\frac{i \omega}{2}~, \\
\beta_+&=& \frac{i \omega}{2}~, \,\,\,\,\,\,\,\,\,\,\,\, \beta_-=\frac{1}{2}-\frac{i \omega}{2}~, \
\end{eqnarray}
and the constants are given by
\begin{equation}
a_{1,2}=\alpha +\beta \pm \frac{i \kappa}{2 r_{+}}~, 
\end{equation}
\begin{equation}
b_{1,2}=\alpha +\beta \mp \frac{i \kappa}{2 r_{+}}~, 
\end{equation}
\begin{equation}
c=1/2+2\alpha~. 
\end{equation}
The general solution of the hypergeometric equation \eqref{hyp} is {\cite{M. Abramowitz}
\begin{equation}
F(z)=C_{1}\,{_2F_1}\left(
a,b,c;z\right) +C_{2}z^{1-c}{_2F_1}\left(
a-c+1,b-c+1,2-c;z\right)~, 
\end{equation}
and it has three regular singular points at $z=0$, $z=1$ and $z = \infty$. Here, $_2F_1(a,b,c;z)$ is a hypergeometric function and $C_1$, $C_2$ are integration constants. Thus, in the vicinity of the horizon $z=0$ and using the property ${_2 F_1}\left(a,b,c;0 \right)=1$, the function $\psi_1(z)$ behaves as
\begin{equation}
\label{Rhorizon2}
\psi _{1}(z) =C_1 e^{\alpha \ln z}+C_2e^{-\alpha \ln z}~.
\end{equation}
Here the first term represents, for $\alpha=\alpha_-$, an ingoing wave and the second an outgoing wave near the black-hole horizon. Imposing the requirement of
only ingoing waves on the event horizon, we fix $C_2=0$. Then, the radial solution can be written as
\begin{equation}\label{horizonsolution}
\psi_1(z)=C_1 e^{\alpha \ln z}(1-z)^\beta {_2}F{_1}(a,b,c;z)=C_1e^{-\frac{i\omega}{2} \ln z}(1-z)^\beta{_2}F{_1}(a,b,c;z)~.
\end{equation}
To implement boundary conditions at infinity ($z=1$), we
apply Kummer's formula
for the hypergeometric function \cite{M. Abramowitz}
\begin{equation}\label{formu}
 {_2}F{_1}(a,b,c;z)=\frac{\Gamma(c)\Gamma(c-a-b)}{\Gamma(c-a)\Gamma(c-b)} {_2}F{_1}(a,b,a+b-c;1-z)+ (1-z)^{c-a-b}\frac{\Gamma(c)\Gamma(a+b-c)}{\Gamma(a)\Gamma(b)} {_2}F{_1}(c-a,c-b,c-a-b+1;1-z)~,
\end{equation}
Taking into consideration the above expression, the radial function~(\ref{horizonsolution}) reads
\begin{eqnarray}
\nonumber \psi_1(z) = && C_1 e^{-\frac{i\omega}{2} \ln z}(1-z)^\beta\frac{\Gamma(c)\Gamma(c-a-b)}{\Gamma(c-a)\Gamma(c-b)}{_2}F{_1}(a,b,a+b-c;1-z)  \\
&&+C_1 e^{-\frac{i\omega}{2}  \ln z}(1-z)^{1/2-\beta}\frac{\Gamma(c)\Gamma(a+b-c)}{\Gamma(a)\Gamma(b)} {_2}F{_1}(c-a,c-b,c-a-b+1;1-z)~,
\end{eqnarray}
and at spatial infinity it can be written as
\begin{equation}\label{R2}\
R_{asymp.}(z) = C_1 (1-z)^\beta \frac{\Gamma(c)\Gamma(c-a-b)}{\Gamma(c-a)\Gamma(c-b)}+ C_1 (1-z)^{1/2-\beta}\frac{\Gamma(c)\Gamma(a+b-c)}{\Gamma(a)\Gamma(b)}~.
\end{equation}
Thus, imposing as a boundary condition that only outgoing waves exist at spatial infinity, we set $c-a=-n$ or $c-b=-n$ for $n=0,1,2,...$.
Therefore, the QNFs of the three-dimensional Lifshitz black hole are given by
\begin{equation}\label{w1}
\omega = \pm \frac{\kappa}{2 r_{+}}-\frac{i}{2}\left(1+2n\right)~.
\end{equation}
The imaginary part of the QNFs is always negative, so that the propagation of a fermionic field is formally stable in this background. The QNFs obtained from Eq. (\ref{radial2}) corresponding to $\psi_2$ yields the same QNFs (\ref{w1}).
In conformal gravity, any metric conformally related to the Lifshitz black hole (\ref{solution}) is also a solution. In particular, the metric:
\begin{equation}
ds^{2} = r^{\eta}\left(- f(r )dt^{2}+\frac{\ell^2}{r^{2}}\frac{dr^{2}}{f ( r )}+r^{2}d\phi ^{2} \right)~, 
\end{equation}
represents a Lifshitz metric with a hyperscaling violating factor $r^{\eta}$, where $\eta$ is an arbitrary exponent.
It can be shown that the QNFs for a massless fermionic field are given by (\ref{w1}) also in this case, which is a consequence of the conformal invariance of the Dirac equation for massless particles, see for instance \cite{Moderski:2008nq}.

On the other hand, following the same procedure above, the quasinormal modes of the soliton metric (\ref{gso}) can be obtained. However, due to the symmetries of the transformations involved, the fermionic quasinormal modes for the soliton can be easily obtained by performing the following substitutions in (\ref{w1})
\begin{equation}
\omega \rightarrow i \kappa_{soliton}, \,\,\, \kappa \rightarrow -i r_{+} \omega_{soliton}~,
\end{equation}
which yields
\begin{equation}
\omega_{soliton}=\pm \left( 2 \kappa_{soliton} +n + \frac{1}{2} \right)~.
\end{equation}
The frequencies are purely real and correspond to normal modes of the fermionic field in the background (\ref{gso}), and thus the fermionic field is stable in this background.

\subsection{Area spectrum}

Bekenstein \cite{Bekenstein:1974jk} was the first to propose the idea that
in quantum gravity the area of black hole horizon is quantized, leading to a
discrete spectrum which is evenly spaced. Then, Hod \cite{Hod:1998vk}
conjectured that the asymptotic QNF is related to the quantized black hole
area, by identifying the vibrational frequency with the real part of the
QNFs. However, it is not universal for every black hole background. Then,
Kunstatter \cite{Kunstatter:2002pj} proposed that the black hole spectrum can
be obtained by imposing the Bohr-Sommerfeld quantization condition to an
adiabatic invariant quantity involving the energy and the vibrational
frequency. Furthermore, Maggiore \cite{Maggiore:2007nq} argued that in the
large damping limit the identification of the vibrational frequency with the
imaginary part of the QNF could lead to the Bekenstein universal bound.

The area and entropy spectrum for the geometry described by (\ref{solution}) was calculated in Ref. \cite{Fernando:2015gta}, employing the QNMs of a conformally coupled scalar field. In this section we obtain the area and entropy spectrum by two methods for the same geometry using the fermionic quasinormal modes obtained in the previous section. The first method we call modified Hod's conjecture, which is based on the ideas by Hod and Maggiore; and the second method we call  modified Kunstatter's conjecture, which is based on the ideas by Kunstatter and Maggiore.

\subsubsection{Modified Hod's conjecture}

Maggiore \cite{Maggiore:2007nq} argued that when a classical black hole is perturbed, its relaxation is governed by a set of QNMs with complex frequencies, whose
behavior is the same as that of damped harmonic oscillators whose real frequencies are
\begin{equation}
\omega_c=\sqrt{|Re(\omega)|^2+|Im(\omega)|^2}~.
\end{equation}
For large values of $n$, $|Im(\omega)| >> Re(\omega)$. So, the transition frequency between adjoining frequencies is given by
\begin{equation}\label{wtt}
\omega_t=\lim_{n \rightarrow \infty} \left( |Im(\omega_{n+1})|-|Im(\omega_{n})| \right)~,
\end{equation}
and the modified Hod's conjecture yields
\begin{equation}
\Delta \mathcal{M}= \hbar \lim _{n \rightarrow \infty} \left( |Im(\omega_{n+1})|-|Im(\omega_{n})| \right) = \hbar~.
\end{equation}
Thus, the mass is quantized and is evenly spaced:
\begin{equation}
\mathcal{M}_n= \hbar n~.
\end{equation}
From this expression it is straightforward to obtain the area spectrum, which is given by
\begin{equation}
\Delta A = 2 \pi \hbar~.
\end{equation}
This same expression was found in \cite{Fernando:2015gta} by employing the QNFs of a conformally coupled scalar field propagating in the Lifshitz black hole.

\subsubsection{Modified Kunstatter's conjecture}
The adiabatic invariant is given by
\begin{equation}
I=\int \frac{d \mathcal{M}}{\omega_t}~,
\end{equation}
and considering the Born-Sommerfeld quantization for a large $n$, we have
\begin{equation}
I \approx n \hbar~,
\end{equation}
where the transition frequency is $\omega_t=1$ according to (\ref{wtt}); thus, $I= \mathcal{M}=n \hbar$, which is the same mass spectrum obtained above by employing the modified Hod's conjecture. Therefore, the area spectrum of this black hole is
\begin{equation} \label{area}
A_n=2 \pi \hbar n~.
\end{equation}
Given that the area spectrum for conformally coupled scalar fields and fermionic fields are the same, it seems that the area spectrum does not depend on the type of perturbations. 
Also, notice that  both methods produce identical values for the area spectrum and it is quantized and equally spaced.

\subsection{Absorption cross section}
\label{coeff}
The reflection and transmission coefficients depend on the behavior
of the radial function both at the horizon and at the asymptotic
infinity, and they are defined by
\begin{equation}\label{reflectiond}\
R =\left|\frac{\mathcal{F}_{\mbox{\tiny asymp}}^{\mbox{\tiny
out}}}{\mathcal{F}_{\mbox{\tiny asymp}}^{\mbox{\tiny in}}}\right|, \,\,\, T=\left|\frac{\mathcal{F}_{\mbox{\tiny
hor}}^{\mbox{\tiny in}}}{\mathcal{F}_{\mbox{\tiny asymp}}^{\mbox{\tiny
in}}}\right|~,
\end{equation}
where $\mathcal{F}$ is the flux. $\mathcal{F}_{hor}^{in}$ denotes the ingoing flux of particles on the event horizon, and $\mathcal{F}_{asymp}^{in}$ and $\mathcal{F}_{asymp}^{out}$ denote the incoming and outgoing flux of particles at spatial infinity, respectively. The flux is given by 
\begin{equation}\label{flux}
\mathcal{F} =\sqrt{-g}\bar{\psi}\gamma ^{r}\psi~, 
\end{equation}
where $\gamma ^{r}=e^{r}_{\,\,1}\gamma ^{1}$, $\bar{\psi}=\psi ^{\dagger
}\gamma ^{0}$,
$\sqrt{-g}=1$, 
and
$e^{r}_{\,\,1}=r \sqrt{f\left(r\right)}$,
which yields
\begin{equation}\label{flux1} 
\mathcal{F}=
\left\vert \psi _{1}\right\vert ^{2}-\left\vert \psi _{2}\right\vert^{2}~.
\end{equation}
The solution for $\psi_1$ is given by (\ref{horizonsolution}):
\begin{equation}\label{s1}
\psi_1=C_1 z^{\alpha} (1-z)^{\beta}  \,\,_2F_1 (a,b,c;z)~,
\end{equation}
and the solution for $\psi_2$ can be obtained from Eq. (\ref{system}), which in terms of the variable $z$ reads
\begin{equation}\label{diff}
\psi'_{2}(z)-\frac{i \omega}{2 z(1-z)} \psi_{2}(z)+\frac{i \kappa}{2r_+  (z(1-z))^{1/2}} \psi_1(z)=0\,.
\end{equation}
In order to solve this equation for $\psi_{2}$, we use the integrating factor $(z/(1-z))^{-i \omega/2}$ and the following property of the hypergeometric function:
\begin{equation}
\int z^{c-1} (1-z)^{a+b-c}\,\,_2F_1(a,b,c;z)=\frac{z^c}{c}\,\,_2F_1(c-a,c-b,c+1;z)~,
\end{equation}
and from (\ref{diff}) we obtain
\begin{equation}\label{s2}
\psi_2 (z)= -C_1 \left(  \frac{z}{1-z} \right)^{i \omega /2} \frac{i \kappa z^c}{2 r_+ c} \,\,_2F_1(c-a,c-b,c+1;z)\,.
\end{equation}
Now, taking into account the behavior of $\psi_1$ and $\psi_{2}$ at the horizon $z \rightarrow 0$, and using~(\ref{flux1}), we get the
flux at the horizon
\begin{equation}
\mathcal{F}
_{hor}^{in}= |C_{1}|^{2}~.
\end{equation}
On the other hand,  by substituting Kummer's formula (\ref{formu}) in (\ref{s1}) and (\ref{s2}) and using Eq. (\ref{flux1}), we obtain the flux at the asymptotic region
\begin{equation}\label{fluxdinfinity}\
\mathcal{F}_{asymp}= \left|C_1\right|^2 \left( \left|A\right|^2-\frac{4r_+^2 (\omega^2+1/4)}{\kappa^2}\left|B\right|^2 \right)~,
\end{equation}
where
\begin{equation}
A  =\frac{\Gamma \left( c\right) \Gamma \left(
c-a-b\right) }{\Gamma \left( c-a\right) \Gamma \left( c-b\right) }~,\,\,\, B  =\frac{\Gamma \left( c\right) \Gamma \left(a+b-c\right) }{\Gamma \left(a\right) \Gamma \left(b\right) }~.
\end{equation}
Therefore, the reflection and transmission coefficients
 are given by
\begin{equation}
R=\frac{4 r_{+}^{2} (\omega^2+1/4)\left|B\right|^2}{ \kappa^2 \left|A\right|^2}
~, \,\,\, T=\frac{1}{\left|A\right|^2}~,\label{coef12}
\end{equation}
and the absorption cross section $\sigma_{abs}$, becomes
\begin{equation}\label{absorptioncrosssection2}\
\sigma_{abs}=\frac{1}{\omega}\frac{1}{|A|^{2}}~.
\end{equation}
Now, we will carry out a numerical analysis of the reflection and transmission coefficients (\ref{coef12}) as well as the absorption cross section~(\ref{absorptioncrosssection2}) of the Lifshitz black hole for fermionic fields.
So, we plot the reflection and transmission coefficients and the absorption cross section in Fig.~(\ref{coefficients}) for $r_{+}=1$ and $\kappa=1$. 
Essentially, we find that the reflection coefficient is greater than the absorption coefficient at the low frequency limit, and at the high frequency limit the reflection coefficient approaches zero, whereas the transmission coefficient tends to $1$. The relation $R+T=1$ is always satisfied, as expected. Note that the transmission coefficient takes a non-null value for $\omega=0$, which is in contrast to what happens for scalar field perturbations \cite{Catalan:2014una}, where the transmission coefficient is always null for $\omega=0$. At the high frequency limit, the behavior for scalar and fermionic fields is similar. Also, we observe that at the low frequency limit, there is a range of values of $\kappa$ which contribute to the absorption cross section (see Fig. \ref{am1}), as occurs in \cite{Gonzalez:2010ht, Gonzalez:2010vv, Gonzalez:2011du, Lepe:2012zf, Gonzalez:2012xc,  Catalan:2014ama, Becar:2014aka} and at the high frequency limit the absorption cross section tends to zero. In addition, we observe a local minimum and a local maximum in the region of low frequencies. On the other hand, the poles of the transmission coefficient coincide with the previously calculated QNFs: $|A|=0$ occurs when $c-a=-n$ or $c-b=-n$, which yields the QNFs given in (\ref{w1}).
\begin{figure}[h]
\begin{center}
\includegraphics[width=0.60\textwidth]{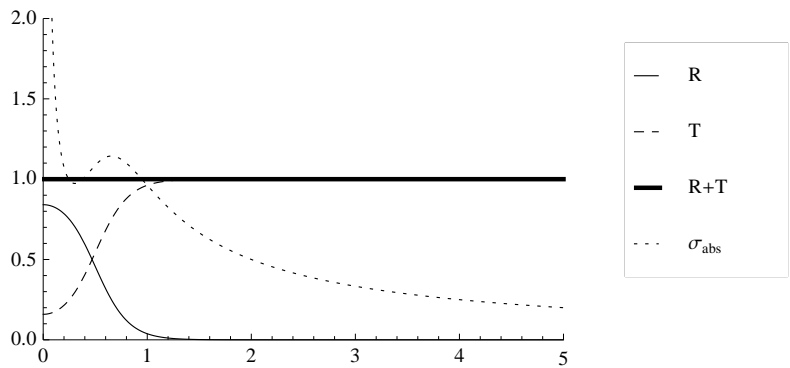}
\caption{The reflection coefficient $R$ (solid curve), the transmission coefficient $T$ (dashed curve),  $R+T$ (thick curve), and the absorption cross section $\sigma_{abs}$ (dotted curve) as a function of $\omega$; for $\kappa=1$ and $r_{+}=1$.}
\label{coefficients}
\end{center}
\end{figure}
\begin{figure}[h]
\begin{center}
\includegraphics[width=0.60\textwidth]{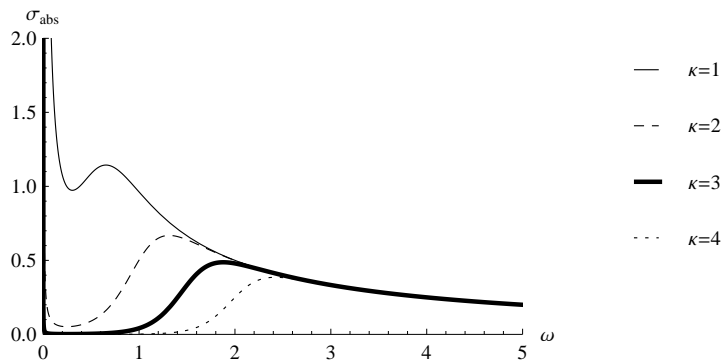}
\caption{The absorption cross section $\sigma_{abs}$ as a function of $\omega$; for $\kappa=1,2,3,4$ and $r_{+}=1$.}
\label{am1}
\end{center}
\end{figure}

\section{Final Remarks}

In this work we studied fermionic field perturbations of an asymptotically Lifshitz black hole in three-dimensional conformal gravity with dynamical exponent $z=0$. In this case, the anisotropic scale invariance corresponds to a space-like scale invariance with no transformation of time. First, we calculated analytically the QNFs of massless fermionic 
perturbations, and we found that the imaginary part of the QNFs is always negative, so that the propagation of a fermionic field is stable in this background. Also, a gravitational soliton solution was obtained by performing the appropriate Wick rotations along with a change of variables to the Lifshitz black hole metric, and the propagation of the fermionic field was shown to be stable on this soliton.
We also obtained the area spectrum of the Lifshitz black hole by means of the application of two methods: the first method we called ``modified Hod's conjecture", and the second ``modified Kunstatter's conjecture", both of which consider the original quantization along with the ideas of Maggiore, and both of which produce identical values for the area spectrum, which is quantized and equally spaced. It is worth mentioning that the area spectrum obtained in \cite{Fernando:2015gta} by employing the QNFs of a conformally coupled scalar field are identical to that obtained here by considering the QNFs for the fermionic fields, with this result suggesting that the area spectrum does not depend on the type of perturbations. So, it would be very interesting to study also the gravitational perturbations in this background and calculate the area spectrum in order to verify whether the same spectrum of the horizon area (\ref{area}) is obtained.
Finally, we calculated the reflection and transmission coefficients and the absorption cross section. We observed that at the low frequency limit, there is a range of values of $\kappa$ which contribute to the absorption cross section and at the high frequency limit the absorption cross section vanishes. 
On the other hand, the reflection coefficient is greater than the transmition coefficient at the low frequency limit and the reflection coefficient is null at high frequencies, while the transmition coefficient tends to 1, and the relation $R+T=1$ is always satisfied. Moreover,  the transmission coefficient takes a non-null value for $\omega=0$, which is in contrast to what happens for scalar field perturbations \cite{Catalan:2014una}. In addition, we showed that the poles of the transmission coefficient coincide with the QNFs for massless fermionic perturbations.

\section*{Acknowledgments}

This work was partially funded by the Comisi\'{o}n
Nacional de Ciencias y Tecnolog\'{i}a through FONDECYT Grants 11140674 (P.A.G. and R.N.V.) and by the Direcci\'{o}n de Investigaci\'{o}n y Desarrollo de la Universidad de La Serena (Y.V. and R.N.V.).
P. A. G. acknowledges the hospitality of the Universidad de La Serena, where part of this work was undertaken.

\appendix

\end{document}